\documentstyle[preprint,floats,epsf,prl,aps]{revtex}
\begin{document}
\preprint{ cond-mat/9407068 }
\title{How do frustrations without disorder result in the spin-glass-like
behavior of the kagom\'{e} antiferromagnet?}
\author{V.B. Cherepanov
}
\address{School of Physics and Astronomy,  Raymond and Beverly Sackler
Faculty of Exact Sciences, Tel~Aviv University, Tel~Aviv, 69978,
Israel and \\
Department of Physics of Complex Systems, Weizmann Institute of
Science, Rehovot, 76100, Israel}

\date{July 13, 1994}
\maketitle
\begin{abstract}
Freezing of the classical spin liquid of the Heisenberg kagom\'{e}
antiferromagnet is studied. At low temperature, the coplanar spin
configurations are known to dominate among all possible
three-dimensional low-energy states because of the fluctuation
interaction.
It is shown that the statistical weight of
${\bf q} =0$ domains is negligible.
This allows one to describe coplanar states in terms of  $\sqrt{3}
 \times \sqrt{3}$ domains with three possible spin orientations.
The domain walls are mapped onto closed self-avoiding loops on the
hexagonal lattice.
The probability of domain formation is evaluated. It is shown that in FC
(field cooled) regime a telescopic hierarchy of domains
appears.
The spatial and temporal behavior of the spin correlation
function is estimated.
The difference between FC and ZFC behavior is discussed.

\end{abstract}



It is usually assumed that the irreversibility
of spin glasses results from interference of frustrations and randomness
\cite{BY,FH}. Nevertheless, the
highly frustrated  Heisenberg antiferromagnets,
with the two-dimensional kagom\'{e} and the three-dimensional pyrochlore
lattices, exhibit the macroscopic properties characteristic for spin
glasses
even when they have ordered structures and non-fluctuating exchange
constants\cite{exp,GRM}.
The main purpose of this paper is to study the
mechanism of the spin-glass-like behavior of the classical
kagom\'{e} antiferromagnet with the nearest neighbor Heisenberg
exchange interaction and to show that
it is due to a
domain hierarchy arising in this magnet at low temperature.

The nearest neighbor exchange interaction is known to dominate in
the kagom\'{e} antiferromagnet \cite{exp,CCR}. The
Hamiltonian   of the
classical kagom\'{e} antiferromagnet  can be represented as a sum of
the total spins in the triangles of the nearest neighbors, ${\bf
S}_{\Delta}$:
\begin{equation}
H = 1/2 \; J \sum_{\Delta} \left( {\bf S}_{\Delta}
\right)^{2},
\end{equation}
where $\Delta$ numbers the triangles of the nearest neighbors. Each spin
participates in two triangles and it contributes into two terms
 ${\bf S}_{\Delta}$. The ground state energy is equal to zero and there
are infinitely many ground states with ${\bf S}_{\Delta} = 0 $.

In antiferromagnetic SCGO, which has been studied experimentally
in some detail, the exchange interaction constant $J \approx 500$ K.
The transition to the spin-glass-like state occurs at $T_{f}
\approx 4$ K, at the temperature two orders lower than the exchange
interaction energy. One could relate this transition to some other
relatively weak interaction which was not clearly indicated
experimentally because of its weakness. However, such a ``mechanism''
for irreversibility unlikely takes place because the numerical
simulations of the model (1) \cite{RB,SH,HR} with no additional
interactions agree with the low temperature experimental results not only
qualitatively but also quantitatively.
In particular, the results of extensive MC simulations of the low
temperature properties of the model (1)
show that the classical kagom\'{e} antiferromagnet with the
nearest neighbor interaction exhibits irreversible behavior
when $T < T_{f}$, $T_{f}/J = 8 \cdot 10^{-3}$ \cite{RB}. The
experimentally measured ratio of the
transition temperature to the exchange interaction energy lies in the
interval $(7.2 - 8.5)\cdot 10^{-3}$ \cite{SP}.
This means that (i) the model (1) possesses the low temperature glass
phase and (ii) the temperature of the experimentally observed
transition is very close to that found in numerical simulations of the
model. Therefore, relatively weak interactions neglected in Eq.~(1)
probably are not related to the spin-glass-like transition.

The classical kagom\'{e} antiferromagnet has a macroscopically
large number of zero energy modes corresponding to continuous transitions
between various ground states \cite{HKB,CHS}. Since any ground state
obeys the condition that the sum of the spins
in each triad of the nearest neighbors is equal to zero, and it is
the only condition for a ground state, one can consider all possible
foldings of the triangular lattice in the three-dimensional spin
space which leave the spin triangles rigid. Any
folded configuration of the triangular lattice represents a
ground state of the kagom\'{e} antiferromagnet \cite{we}. It can be
shown that
there exists one-to-one correspondence between the ground states
and the foldings of the triangular lattice (``origami'') in the
3D spin space.
The zero energy modes
correspond to the foldings.

The statistical weight of thermally excited states
near coplanar ground states dominates over the
weight of all other low energy states \cite{CHS}.
This means that the coplanar states and the
near-coplanar states have the largest entropy because these states are
most flexible states of the triangular tethered membrane.
Now we introduce a geometrical classification of the coplanar ground states.
 Marking the axis of foldings on the triangular lattice yields a
prescription how to get from the state ${\bf q} =0$ to any folded state.
Each
line depicts a boundary between different domains with ${\bf q} =0$ order.
Thus, the map of folding lines on the triangular lattice provides
a geometrical representation of the coplanar states in terms of boundaries of
${\bf q} =0$ domains on the plane. This map consists
of a number of meetings and intersections of straight lines. There are only
four possible configurations of the line meetings and intersections. They
are shown in the fig.~1 as continuous lines:
(a) meeting of four lines (``fork''), (b) straight line,
 (c) six radii star, and
(d) no lines at all, it is  the homogeneous state ${\bf q} =0$. The axis
of folding depict the boundaries between domains with ${\bf q} =0$
order.
\begin{figure}
\epsfxsize = \hsize
\epsffile{kfig.eps}

\caption{Elements of the domain boundaries. (a - d) -- boundaries
separating {\bf q} = 0 domains (continuous lines), (e - h) --
dual domain boundaries. Continuous
lines in (a - d) correspond to dashed lines in (e - h) and vice versa.
Dashed lines represent no folding/unfolding operations. }
\end{figure}
The state $ \sqrt{3} \times \sqrt{3}$ results from folding of the
planar triangular lattice into one triangle, its folding map consists
of the six radii stars located in each triangle junction, i.~e. it
is the whole triangular lattice of the folding axis. I will show
further that the statistical weight of the state
 $ \sqrt{3} \times \sqrt{3}$ and the states close to it is much
larger than that of the state ${\bf q} =0$. In order to describe the
states close to $ \sqrt{3} \times \sqrt{3}$, another geometrical
representation of the ground states, dual to the folding line
map described above, is more convenient. Each coplanar state can be
obtained by
a sequence of $180^{\circ}$ unfoldings of the stack of the spin
triangles which represents the state  $ \sqrt{3} \times
\sqrt{3}$. If one depicts the
axis of unfolding one obtains another geometrical representation of
the coplanar states. There are four possible elements that form the
dual map shown in fig.~1:
(e) 120-degree angle, (f) cross constructed of two angles (fig.~1(e)),
(g) the state  $ \sqrt{3} \times \sqrt{3}$ with no unfolding lines, and
 (h) six radii star. These lines correspond to boundaries between
domains of different states $ \sqrt{3} \times \sqrt{3}$. There are
three kinds of domain orientations arising from unfolding along three axis
on the triangular lattice. In the dual representation, a ${\bf q} =0$
domain is covered with the six radii stars, the smallest ${\bf q} =0$
domain is represented by one star (h).

The simplest spin configuration close to the homogeneous
state $ \sqrt{3} \times \sqrt{3}$, the weather vane defect,
is produced by rotation of a group of six neighboring spin
triangles. There are three different ways to rotate the triangles
along each of the triangular lattice axis. If $ \theta_{1},
  \theta_{2}$, and $\theta_{3}$ are the the angles of those rotations
the energy of a low-energy configuration for the small angles takes
the form:
$ V( \theta_{1}, \theta_{2}, \theta_{3}) =
\alpha_{0} J  \left( \theta_{1}^{2} \theta_{2}^{2} +
 \theta_{2}^{2} \theta_{3}^{2} + \theta_{3}^{2} \theta_{1}^{2} \right) $
where $\alpha_{0} $ is a dimensionless coefficient. At low temperature,
only the states in the vicinity of the coplanar state with
$ | \theta_{i} | \lesssim (T/\alpha_{0}J)^{1/4} $ contribute to the
statistical sum \cite{CHS}.

The energy of a state in the vicinity of an arbitrary coplanar
ground state is
$ V( \theta_{i} ) = J \sum_{i,j} \alpha_{i,j}
 \theta_{i}^{2} \theta_{j}^{2}. $
For the state $ \sqrt{3} \times \sqrt{3}$ the coefficients
$ \alpha_{i,j} $ are equal to zero for most pairs of the angles
$ ( \theta_{i}, \; \theta_{j} ) $. They are nonzero for pairs of
angles involved in the same weather-vane defects only.
In the state ${\bf q} =0$, all the angles inside the domain, except foldings
along parallel axis, are involved,
i. e. $\alpha_{i,j} = \alpha_{1}  $ for the majority of the pairs.

Comparing the statistical weights of the states near
 coplanar  $ \sqrt{3} \times \sqrt{3}$ and ${\bf q} =0$ configurations
one can find that
the statistical sum $ Z_{{\bf q}=0} $ rapidly decreases with the number
of spins $N$ and
\begin{equation}
\frac{Z_{{\bf q}=0}}{ Z_{\sqrt{3} \times \sqrt{3}}} \approx
\left(\frac{\alpha_{0}}{\alpha_{1}} \right) ^{N/4} \frac{
\Gamma{(N/4)} /
\Gamma{(N/2)} }{\left[ \Gamma{(3/4)} / \Gamma{(3/2)}
\right] ^{N/3}}.
\end{equation}
The smallest size of a ${\bf q} =0$ domain is $N = 6 $.
In addition, such domains
appear in pairs, therefore the statistical weight of ${\bf q} =0$ domains
is strongly suppressed in comparison with that of $ \sqrt{3} \times \sqrt{3}$
domains. The weight a pair of domains ${\bf q} =0$ is  $\epsilon \approx
2^{-12} (\alpha_{0}/\alpha_{1} )^{3}$ times smaller than
the weight of the $ \sqrt{3} \times \sqrt{3}$ state.
 An analysis of results of MC simulations
\cite{RB} shows that there was only one pair of smallest possible
${\bf q} =0$ domains, with $N=12$, on the lattice with 576  spins observed.

The small parameter $\epsilon $ allows one to
neglect ${\bf q} =0$ domains. This provides a possibility for a simple
geometrical classification of the domain boundaries:
the domain boundaries form closed loops consisting of $120^{\circ}$
elements (fig.~1 (e)), and if we have two loops, only two alternatives
exist:
either one loop is entirely inside another loop or the loops are outside
each other.

Now we evaluate the probability of a domain
formation against the homogeneous  $ \sqrt{3} \times \sqrt{3}$ background
and the probability of dissipation of such a domain. This case corresponds
to FC conditions because
the thermal fluctuations select the state
 $ \sqrt{3} \times \sqrt{3}$  in a
finite magnetic field \cite{SH}.

Since no ground state dynamics may exist
we employ the relaxation
dynamics formalism\cite{FH,ZJ} in order to describe the domain kinetics
under influence of thermal fluctuations. We start with the Langevin
equation for the angles $\theta_{i}$ and
consider the case when a group of many
spin triangles
is rotated as a whole domain. In this case, only one  $180^{\circ }$
rotation is necessary to overcome the free energy barrier between
the initial and final coplanar configurations.
(The case of a single
weather vane defect rotation was considered in \cite{DH}).
The
potential of low energy states in the vicinity of such a trajectory is:
\begin{equation}
V\{ \theta_{i} \} =  J  \sum_{j=1}^{L} f(\theta_{0})
\theta_{j}^{2},
\end{equation}
where the angle $\theta_{0}$ goes from $0$ to $\pi $, it corresponds
to the domain rotation, and the angles $\theta_{j}$ describe small
fluctuations of the spin triangles in all other directions. The
function $f(\theta_{0})$ is periodic, its expansion in series of
$\theta_{0}$ starts from quadratic terms at both $\theta_{0} = 0$
and  $\theta_{0} = \pi $.

The probability for the system to
reach the configuration $\{ \theta_{i}^{(1)} \}$ at the moment $\tau_{1}$
starting from the configuration $\{ \theta_{i}^{(0)} \}$ at the moment
$\tau_{0}$ is
\begin{eqnarray}
P\{\theta_{i} \} = \int \exp{ \left( -
\int_{\tau_{0}}^{\tau_{1}} {\cal L}\{\theta_{i}(\tau ) \} d\tau \right) }
D\theta_{i}(\tau ),   \nonumber \\
{\cal L} \{\theta_{i}(\tau ) \} = 1/2 \; (\partial \theta_{0} / \partial \tau
)^{2} +  1/2 \sum_{j=1}^{L} (\partial \theta_{j} / \partial \tau
)^{2} \nonumber \\
+ 1/2 \; (\beta J)^{2}\left( f^{2}(\theta_{0}) +
f'^{2}(\theta_{0}) \sum_{j} {\theta_{j}}^{2} \right)
\sum_{j} {\theta_{j}}^{2} \nonumber \\
+ 1/2 \; (\beta J) {f''(\theta_{0})}  \sum_{j} {\theta_{j}}^{2}
+ 1/2 \; L (\beta J) {f(\theta_{0})}
\end{eqnarray}
where $\beta = 1/T $ and $ \tau $ is the dimensionless time. The
probability to get from a
state in the vicinity of the coplanar ground state $
\theta_{0}^{(0)} =0 $ to the states near $ \theta_{0}^{(1)} = \pi $ is the
product of the probability to overcome the free energy barrier between the
initial and the final states and the statistical weight
of the final state. At finite $\theta_{0}$, the
$\theta_{j}$ degrees of freedom are stiff. After integrating over
these fast degrees of freedom one obtains an effective potential for
the slow rotation $\theta_{0}$. This effective potential is
proportional to the domain perimeter $L$ in the large $L$ limit. This
results in the effective action $\cal A$ along the trajectory
proportional to
$ \sqrt L$. The probability to overcome the barrier along the instanton
path is proportional to $\exp{({\cal -A})}$.
The ratio of the statistical weights of final and initial states
 $ W_{L} $
can be estimated as follows:
\begin{equation}
W_{L} = \frac{\int \Pi d \theta_{i} \times \nonumber \\
\exp{\left[ - (\beta J )
\left( \alpha_{1}{ \sum^{L}_{j=1}} \theta_{0}^{2}
\theta_{j}^{2}
\right) \right]}}{
\int \Pi d \theta_{i} \times \nonumber \\
\exp{\left[ - (\beta J )
\left( \alpha_{0}{ \sum^{L}_{j=1}} \theta_{0}^{2}
\theta_{j}^{2}
\right) \right]}}
\label{W_L}
\end{equation}
where
summation runs over the rotating triangles.
If all the angles $ \theta_{j} $ in the numerator (\ref{W_L})
were independent, the ratio $W_{L}$ (\ref{W_L}) would be equal to
$ W_{L} =  \exp{ \left[ - 1/2 \;
\ln{\left( \alpha_{0} /\alpha_{1} \right)} L \right]}.$
Actually, not all
the angles  $ \theta_{j} $ are independent.
Nevertheless, it can be shown that
\begin{equation}
W_{L} \propto \exp{ (- \rho L )}, \; \; \; \; \rho
\sim 1
\end{equation}
in the $L \rightarrow \infty $ limit.

The probability of formation of any domain with the perimeter
length $L$ is equal to the product of the probability for a single
domain to appear, $W_{L}$, and the number of domains with the boundary
length $L$, $N_{L}$. The latter coincides with the number of closed
self-avoiding chains on the hexagonal lattice \cite{CJ}
\begin{equation}
N_{L} \propto \exp {( \eta L )}, \; \; \; \eta \approx 0.609.
\end{equation}

We arrive at an alternative: if $ \rho < \eta $ then a large domain
arises against a homogeneous  $ \sqrt{3} \times \sqrt{3}$ background
most probably. In the opposite case, $ \rho > \eta $, smallest domains
arise first, they cover all the lattice and prevent from growth of large
domains.

A hierarchical structure of the phase space is a key component to
the slow relaxation and glassy behavior\cite{PSAA}. In the case of
the field cooled kagom\'{e} antiferromagnet, the hierarchy results
from the instability of the homogeneous  $ \sqrt{3} \times \sqrt{3}$
state against domain formation. If $ \rho < \eta $ then
the largest domains are the  most unstable ones. The instability
leads to formation of the largest possible domain, of order of the
sample size, with the perimeter length $ L_{1}$.
All the area inside that domain is a
homogeneous state $ \sqrt{3} \times \sqrt{3}$, the area outside the
domain is also a homogeneous state $ \sqrt{3} \times \sqrt{3}$ of
another orientation. The areas inside and outside the first
domain are unstable with respect to domain formation. Let us
concentrate on domain formation inside the first domain.
If one waits for some time, a smaller
domain inside the first one appears. The perimeter of the second
domain is $ L_{2} \approx L_{1} - \Delta L $, where
$\Delta L = \max \left[ (\eta-\rho)^{-1}, 12 \right] $.
If one waits for a long time
one observes formation of smaller  and smaller domains inside larger
ones. In the limit of large $L$, only one domain of the size $ \approx
L - \Delta L $ appears inside the domain of the previous generation, with
the perimeter $L$.
The hierarchy in such a domain structure is
telescopic: a smaller domain is subordinated to a larger one whose
boundary surrounds the former.
 In the opposite case, $ \rho > \eta $, no hierarchy arises and no
irreversibility exists. The experiment \cite{exp} and especially MC
simulations \cite{RB}, where the telescopic domain hierarchy can be seen
in a snap-shot picture, indicate that $\rho < \eta $ in the kagom\'{e}
antiferromagnet.

Now we estimate the spatial and temporal dependencies of
the spin correlation function. The spatial correlation
function $C_{\sqrt{3} \times \sqrt{3}}({\bf r,r'})$
corresponding to the $\sqrt{3} \times \sqrt{3}$ order
\cite{RB}. If there are domains in a sample the
correlation
function can be equal either to 1 if the spins belong to domains with
the same orientation, or to $ - 1/2$ if the spins belong to domains with
different orientations, and there are two possible choices of
different domain orientations.
In the sample filled with domains
the correlation function $C_{\sqrt{3} \times \sqrt{3}}({\bf r,r'})$ is
equal to 1 if there is a path connecting the spins that does not cross
any domain boundary. If the spin $ {\bf S}({\bf r})$ belongs to a
domain of the size $L$, the correlation function can be estimated as
follows: it is $ \propto 1/|{\bf r - r'}|$ if $|{\bf r - r'}| < L $
and it is equal to zero if  $|{\bf r - r'}| \gg L $. Averaging over the
spin positions ${\bf r}$ yields  the estimate
\begin{equation}
C_{\sqrt{3} \times \sqrt{3}}({\bf r,r'}) \propto 1/|{\bf r - r'}|
\end{equation}
This estimate is in a good agreement with the results of MC simulations
\cite{RB} that $C_{\sqrt{3} \times
\sqrt{3}}({\bf r,r'}) \propto |{\bf r - r'}|^{-0.93}$ in the case of
relaxation starting with the initial homogeneous state $\sqrt{3}
\times \sqrt{3}$.

Let us estimate the time dependence of the spin correlation function
$ C(t) = \langle {\bf S}({\bf r},0) {\bf S}({\bf r},t)
\rangle $. If the spin $ {\bf S(r)}$ belongs to a domain of the size
$L$, it rotates with a characteristic period $ t_{L}$. The
product ${\bf S}({\bf r},t_{0}) {\bf S}({\bf r},t_{0} + \Delta t)$
changes its sign with the characteristic time $ \Delta t \sim t_{L}$.
Therefore
\begin{equation}
\frac{\Delta C(t)}{\Delta t} \propto
- \frac{C(t)}{N} \frac{A_{L + \Delta L} {\cal N}_{L + \Delta L}
- A_{L}{\cal N}_{L} }{\Delta L} \frac{ \Delta L}{\Delta t}.
\end{equation}
Here $N$ is the number of the spins in the sample,
 the number of spins in a domain of the size $L$ is
proportional to its area $A_{L}$, ${\cal N}_{L}$ is the number of
domains of the size $L$. The size of the domains $L$ is connected with
time $t$:
$ t_{L} = w_{0}^{-1} \exp ({\alpha L})$.
In general, one can expect that
\begin{equation}
A_{L} \propto L^{2 \nu}, \; \;
{\cal N}_{L} \propto L^{\zeta}.
\end{equation}
This results in in the following time decay of the spin correlation
function:
\begin{equation}
\ln \left(\frac{C({\bf r}, t)}{ C({\bf r}, 0)} \right)
\propto - \left[ \ln \left( w_{0} t \right)
\right]^{\zeta + 2 \nu}.
\end{equation}
In the case of FC regime, we have the telescopic domain hierarchy with
$\zeta = 0$. The exponent $2 \nu $ is the mean area of a random
self-avoiding loop
with the perimeter $L$ on the hexagonal lattice. The exponent $\nu $
is bounded by the exponent for the mean squared distance between
elements of a random self-avoiding chain on the hexagonal lattice,
$\nu_{low} = 3/4 $ \cite{CJ}, and $\nu_{high}=1$.
Thus the spin correlation exhibits  a  stretched exponential
relaxation characteristic
for glasses.

In the ZFC regime, the relaxation starts from a random domain
distribution, with the exponent $\zeta \approx - 4/3 $ \cite{HR,RCC}. The
small domains do not allow new large domains to appear because of the
 no boundary crossing constraint. Qualitatively, this results in a
faster decay of the spin correlation function $ C_{\sqrt{3} \times
\sqrt{3}}({\bf r}) $ with the distance and
in a different susceptibility in comparison with the case of the FC
regime.

In conclusion, it has been shown that: (i) The local $ \sqrt{3} \times
\sqrt{3}$ order dominates at low temperature, this explains results of
experimental observations \cite{exp,BAEC}. (ii) In the field cooled regime a
telescopic hierarchy of domains arises. This hierarchy leads to slow
spin relaxation, the estimates of the correlation function decay
agree with results of MC simulations \cite{RB}. The reason for
the difference between FC and ZFC spin correlations is justified. Thus,
the origin of the spin-glass-like behavior of the pure kagom\'{e}
antiferromagnet has been clarified.

\acknowledgments

I acknowledge numerous discussions with E. Shender on the initial
stage of the research.
I am grateful
to A.~Aharony, J.~Berlinsky, B.~Botvinnik, M.~Chertkov,
P.~Coleman,
V.~Elser, C.~Henley, D.~Huber, D.~Huse,
A.~Ruckenstein, M.~Schwartz, and P.~Young
for useful discussions.
I am grateful to J.~Reimers and J.~Berlinsky, D.~Huse,
R.~Chandra and P.~Coleman, and E.~Shender and P.~Holdsworth for sending me
preprints of their papers.  I also acknowledge financial support from the
US --- Israel Binational Science Foundation (GIF).

\end{document}